# Thermal Transport in Graphene and Graphene Multilayers

Alexander A. Balandin[1,*] and Denis L. Nika[1,2]

[1]Department of Electrical Engineering and Materials Science and Engineering Program, Bourns College of Engineering, University of California, Riverside, CA 92521 U.S.A.

[2]E. Pokatilov Laboratory of Physics and Engineering of Nanomaterials, Department of Theoretical Physics, Moldova State University, Chisinau, MD-2009, Moldova

## Abstract

In this paper we review thermal properties of graphene and multilayer graphene and discuss the optothermal technique developed for the thermal conductivity measurements. We also outline different theoretical approaches used for the description of phonon transport in graphene and provide comparison with available experimental thermal conductivity data.

# Introduction

Heat removal has become a crucial issue for continuing progress in electronic industry owing to increased levels of dissipated power density and speed of electronic circuits [1]. Self-heating is a major problem in optoelectronics and photonics [2]. These facts stimulated recent interest in thermal properties of materials. Acoustic phonons – fast moving quanta of the crystal lattice vibrations – are the main heat carriers in a variety of material systems. The phonon and thermal properties of nanostructures are substantially different from those of bulk crystals [3-16]. Semiconductor thin films or nanowires do not conduct heat as well as bulk crystals due to increased phonon - boundary scattering [4-5] as well as changes in the phonon dispersion and density of states (DOS) [3-10]. However, theoretical studies suggested that phonon transport in strictly two-dimensional (2D) and one-dimensional (1D) systems can reveal exotic behavior, leading to infinitely large *intrinsic* thermal conductivity [11-12]. These theoretical results have led to discussions of the validity of Fourier's law in low-dimensional systems [17-18] and further stimulated interest in the acoustic phonon transport in 2D systems.

In this chapter, we focus on the specifics of the acoustic phonon transport in graphene. After a brief summary of the basics of thermal physics in nanostructures and experimental data for graphene's thermal conductivity, we discuss, in more detail, various theoretical approaches to calculation of the phonon thermal conductivity in graphene.

### I. Basics of Phonon Transport and Thermal Conductivity

The main experimental technique for investigation of the acoustic phonon transport in a given material system is the measurement of its lattice thermal conductivity [19-20]. In this section, we define the main characteristics of heat conduction. The thermal conductivity is introduced through Fourier's law [21-22]:

$$\vec{\phi} = -K\nabla T, \tag{1}$$

where $\vec{\phi}$ is the heat flux, $\nabla T$ is the temperature gradient and $K = (K_{\alpha\beta})$ is the thermal conductivity tensor. In the isotropic medium, thermal conductivity does not depend on the direction of the heat flow and $K$ is treated as a constant. The latter is valid for the small temperature variations only. In a wide temperature range, thermal conductivity is a function of temperature, i.e. $K \equiv K(T)$. In general, in solid materials heat is carried by phonons and electrons so that $K=K_p+K_e$, where $K_p$ and $K_e$ are the phonon and electron contributions, respectively. In metals or degenerately-doped semiconductors, $K_e$ is dominant due to the large density of free carriers. The value of $K_e$ can be determined from the measurement of the electrical conductivity $\sigma$ via the Wiedemann-Franz law [23]:

$$\frac{K_e}{\sigma T} = \frac{\pi^2 k_B^2}{3e^2}, \qquad (2)$$

where $k_B$ is the Boltzmann's constant and $e$ is the charge of an electron. Phonons are usually the main heat carriers in carbon materials. Even in graphite, which has metal-like properties [24], the heat conduction is dominated by acoustic phonons [25]. This fact is explained by the strong covalent $sp^2$ bonding, resulting in high in-plane phonon group velocities and low crystal lattice unharmonicity for in-plane vibrations.

The phonon thermal conductivity can be written as

$$K_p = \Sigma_j \int C_j(\omega)\upsilon_{x,j}(\omega)\upsilon_{x,j}(\omega)\tau_j(\omega)d\omega, \qquad (3)$$

where summation is performed over the phonon polarization branches $j$, which include two transverse acoustic branches and one longitudinal acoustic branch, $\upsilon_{x,j}$ is the projection of the phonon group velocity $\vec{\upsilon}_j = d\omega_j/d\vec{q}$ on the $X$-axis for the $j$th branch, which, in many solids, can be approximated by the sound velocity, $\tau_j$ is the phonon relaxation time, $C_j = \hbar\omega_j \partial N_0(\hbar\omega_j/k_BT)/\partial T$ is the contribution to heat capacity from the $j$th branch, and $N_0(\frac{\hbar\omega_j}{k_BT}) = [\exp(\frac{\hbar\omega_j}{k_BT})-1]^{-1}$ is the Bose-Einstein phonon equilibrium distribution function. The phonon mean-free path (MFP) $\Lambda$ is related to the relaxation time through the expression $\Lambda = \tau\upsilon$. In the relaxation-time approximation (RTA), various scattering mechanisms, which limit the MFP, usually considered as additive, i.e. $\tau_j^{-1} = \sum_i \tau_{i,j}^{-1}$, where $i$ denotes scattering mechanisms. In typical solids, acoustic phonons, which carry the bulk of heat, are scattered by other phonons, lattice defects, impurities, conduction electrons, and interfaces [26-29].

In ideal crystals, i.e. crystals without lattice defects or rough boundaries, $\Lambda$ is limited by the phonon - phonon scattering due to the crystal lattice anharmonicity. In this case, thermal conductivity is referred to as intrinsic. The anharmonic phonon interactions, which lead to the finite thermal conductivity in three dimensions, can be described by the Umklapp processes [26]. The Umklapp scattering rates depend on the Gruneisen parameter $\gamma$, which determines the degree of the lattice anharmonicity [26-27]. Thermal conductivity is extrinsic when it is mostly limited by the extrinsic effects such phonon – rough boundary or phonon – defect scattering.

In nanostructures, the phonon energy spectra are quantized due to the spatial confinement of the acoustic phonons. The quantization of the phonon energy spectra usually leads to decreasing phonon group velocity. The modification of the phonon energies, group velocities and density of states, together with phonon scattering from boundaries affect the thermal conductivity of nanostructures. In most of cases, the spatial confinement of acoustic phonons results in a reduction of the phonon thermal conductivity [30-31]. However, it was predicted that the thermal conductivity of nanostructures embedded within the acoustically hard barrier layers can be increased via spatial confinement of acoustic phonons [6-7, 10, 32].

The phonon boundary scattering can be evaluated as [29]

$$\frac{1}{\tau_{B,j}} = \frac{\upsilon_{x,j}}{D}\frac{1-p}{1+p}, \tag{4}$$

where $D$ is the nanostructure or grain size and $p$ is the specularity parameter defined as a probability of specular scattering at the boundary. The momentum-conserving specular scattering ($p=1$) does not add to thermal resistance. Only diffuse phonon scattering from rough interfaces ($p\rightarrow 0$), which changes the phonon momentum, limits the phonon MFP. One can find $p$ from the surface roughness or use it as a fitting parameter to experimental data. The commonly used expression for the phonon specularity is given by [29, 33-34]

$$p(\lambda) = \exp(-\frac{16\pi^2\eta^2}{\lambda^2}), \tag{5}$$

where $\eta$ is the root mean square deviation of the height of the surface from the reference plane and $\lambda$ is the wavelength of the incident phonon.

In the case when the phonon - boundary scattering is dominant, thermal conductivity scales with the nanostructure or grain size D as $K_p \sim C_p \upsilon \Lambda \sim C_p \upsilon^2 \tau_B \sim C_p \upsilon D$. In the very small structures with $D<<\Lambda$, the thermal conductivity dependence on the physical size of the structure becomes more complicated due to the strong quantization of the phonon energy spectra [6, 30, 32]. The specific heat $C_p$ depends on the phonon density of states, which leads to different $C_p(T)$ dependences in three-dimensional (3D), two-dimensional and one-dimensional systems, and reflected in $K(T)$ dependence at low $T$ [26, 29]. In bulk at low $T$, $K(T) \sim T^3$ while it is $K(T) \sim T^2$ in 2D systems.

Thermal conductivity $K$ defines how well a given material conducts heat. Another characteristic – thermal diffusivity, $\alpha$ – defines how fast the material conducts heat. Thermal diffusivity is given by the expression

$$\alpha = \frac{K}{C_p \rho_m}, \tag{6}$$

where $\rho_m$ is the mass density. Many experimental techniques measure thermal diffusivity rather than thermal conductivity.

## II. Experimental Data for Thermal Conductivity of Graphene

We start by providing a brief summary of the experimental data available for the thermal conductivity of graphene. The first measurements of heat conduction in graphene [35-40] were carried out at UC Riverside in 2007 (see figure 1). The investigation of the phonon transport was made possible by the development of the optothermal Raman measurement technique. Balandin and co-workers [35-36] took advantage of the fact that graphene has distinctive signatures in Raman spectra with clear *G* peak and *2D* band [41-45]. Moreover, they also found that the *G* peak of graphene's Raman spectra exhibits strong temperature dependence [41]. The latter means that the shift in the position of *G* peak in response to the laser heating can be used for measuring the local temperature rise. The correlation between the temperature rise and amount of power dissipated in graphene, for the sample with given geometry and proper heat sinks, can give the value of the thermal conductivity K (see the schematic of the experiment in Figure 1 (a)). Even a small amount of power dissipated in graphene can be sufficient for inducing a measurable shift in the *G* peak position due to the extremely small thickness of the material – one atomic layer. The suspended portion of graphene served several essential functions for (i) accurately determining the amount of power absorbed by graphene through the calibration procedure; (ii) forming two-dimensional in-plane heat front propagating toward the heat sinks; (iii) and reducing the thermal coupling to the substrate through the increased micro- and nanoscale corrugations (see Figure 1 (b)).

**[Figure 1]**

The long graphene flakes for these measurements were produced using the standard technique of mechanical exfoliation of bulk Kish and highly oriented pyrolytic graphite (HOPG) [46-48]. The trenches were fabricated using the reactive ion etching. The width of these trenches ranged from 1 μm to 5 μm with the nominal depth of 300 nm. The single layer graphene flakes were selected using the micro Raman spectroscopy by checking the intensity ratio of *G* and *2D* peaks and by *2D* band deconvolution [43-45]. The combination of these two Raman techniques with the atomic force microscopy (AFM) and scanning electron microscopy (SEM) allowed authors [35-36] to verify the number of atomic planes and flake uniformity with a high degree of accuracy. It was found that the thermal conductivity varies in a wide range and can exceed that of the bulk graphite, which is ~2000 W/mK at room temperature (RT). It was also determined that the electronic contribution to heat conduction in the un-gated graphene near RT is much smaller than that of phonons, i.e. $K_e<<K_p$. The phonon MFP in graphene was estimated to be on the order of 800 nm near RT [36].

**[Figure 2]**

Several independent studies, which followed, also utilized the Raman optothermal technique but modified it via addition of a power meter under the suspended portion of graphene. It was found that the thermal conductivity of suspended high-quality chemical vapour deposited (CVD) graphene exceeded ~2500 W/mK at 350 K, and it was as high as $K≈1400$ W/mK at 500 K [49]. The reported value was also larger than the thermal conductivity of bulk graphite at RT. Another Raman optothermal study with the suspended graphene found the thermal conductivity in the range from ~1500 to ~5000 W/mK [50]. Another group that repeated the Raman-based measurements found $K≈630$ W/mK for a suspended graphene membrane [51]. The differences in the actual temperature of graphene under laser heating, strain distribution in the suspended graphene of various sizes and geometries can explain the data variation.

Another experimental study reported the thermal conductivity of graphene to be ~1800 W/mK at 325 K and ~710 W/mK at 500 K [52]. These values are lower than that of bulk graphite. However, instead of measuring the light absorption in graphene under conditions of their experiment, the authors of Ref. [52] assumed that the optical absorption coefficient should be 2.3%. It is known that due to many-body effects, the absorption in graphene is the function of wavelength $\lambda$, when $\lambda$>1 eV [53-55]. The absorption of 2.3% is observed only in the near-infrared at ~1 eV. The absorption steadily increases with decreasing $\lambda$ (increasing energy). The 514.5-nm and 488-nm Raman laser lines correspond to 2.41 eV and 2.54 eV, respectively. At 2.41 eV the absorption is about 1.5 ×2.3% ≈ 3.45% [54]. The value of 3.45% is in agreement with the one reported in another independent study [56]. Replacing the assumed 2.3% with 3.45% in the study reported in Ref. [52] gives ~2700 W/mK at 325 K and 1065 W/mK near 500 K. These values are higher than those for the bulk graphite and consistent with the data reported by other groups [49, 56], where the measurements were conducted by the same Raman optothermal technique but with the measured light absorption.

The data for suspended or partially suspended graphene is closer to the intrinsic thermal conductivity because suspension reduces thermal coupling to the substrate and scattering on the substrate defects and impurities. The thermal conductivity of fully supported graphene is smaller. The measurements for exfoliated graphene on $SiO_2$/Si revealed in-plane $K$≈600 W/mK near RT [57]. Solving the Boltzmann transport equation (BTE) and comparing with their experiments, the authors determined that the thermal conductivity of free graphene should be ~3000 W/mK near RT.

Despite the noted data scatter in the reported experimental values of the thermal conductivity of graphene, one can conclude that it is very large compared to that for bulk silicon ($K$=145 W/mK at RT) or bulk copper ($K$=400 W/mK at RT) – important materials for electronic applications. The differences in $K$ of graphene can be attributed to variations in the graphene sample lateral sizes (length and width), thickness non-uniformity due to the mixing between single-layer and few-layer graphene, material quality (e.g. defect concentration and surface contaminations), grain size and orientation, as well as strain distributions. Often the reported thermal conductivity values of graphene corresponded to different sample temperatures $T$, despite the fact that the measurements were conducted at ambient temperature. The strong heating of the samples was required due to the limited spectral resolution of the Raman spectrometers used for temperature measurements. Naturally, the thermal conductivity values determined at ambient but for the samples heated to *T*~350 K and *T*~600 K over a substantial portion of their area would be different and cannot be directly compared. One should also note that the data scatter for thermal conductivity of carbon nanotubes (CNTs) is much larger than that for graphene. For a more detail analysis of the experimental uncertainties the readers are referred to a comprehensive reviews [16,58].

**III. Phonon Transport in Suspended Few-Layer Graphene**

The phonon thermal conductivity undergoes an interesting evolution when the system dimensionality changes from 2D to 3D. This evolution can be studied with the help of suspended few-layer graphene (FLG) with increasing thickness *H* – number of atomic planes *n*. It was reported in Ref. [38] that thermal conductivity of suspended uncapped FLG decreases with increasing *n* approaching the bulk graphite limit (see figure 3). This trend was explained by considering the intrinsic quasi-2D crystal properties described by the phonon Umklapp scattering [38]. As *n* in FLG increases – the phonon dispersion changes and more phase-space states become available for phonon scattering leading to thermal conductivity decrease. The phonon scattering from the top and bottom boundaries in suspended FLG is limited if constant *n* is maintained over the layer length. The small thickness of FLG (*n*<4) also means that phonons do not have transverse cross-plane component in their group velocity leading to even weaker boundary scattering term for the phonons. In thicker FLG films the boundary scattering can increase due to the non-zero cross-plane phonon velocity component. It is also harder to maintain the constant thickness through the whole area of FLG flake. These factors can lead to a thermal conductivity below the graphite limit. The graphite value is recovered for thicker films.

[Figure 3]

The experimentally observed evolution of the thermal conductivity in FLG with *n* varying from 1 to *n*~4 [38] is in agreement with the theory for the crystal lattices described by the Fermi-Pasta-Ulam Hamiltonians [59]. The molecular-dynamics (MD) calculations for graphene nanoribbons with the number of planes *n* from 1 to 8 [60] also gave the thickness dependence of the thermal conductivity in agreement with the UC Riverside experiments [38]. The strong reduction of the thermal conductivity as *n* changes from 1 to 2 is in line with the earlier theoretical predictions [61]. In another reported study, the Boltzmann transport equation was solved under the assumptions that in-plane interactions are described by Tersoff potential while the Lennard-Jones potential models interactions between atoms belonging to different layers [62-63]. The obtained results suggested a strong thermal conductivity decrease as *n* changed from 1 to 2 and slower decrease for *n*>2.

The thermal conductivity dependence on the FLG is entirely different for the encased FLG where thermal transport is limited by the acoustic phonon scattering from the top and bottom boundaries and disorder. The latter is common when FLG is embedded between two layers of dielectrics. An experimental study [64] found $K\approx 160$ W/mK for encased single-layer graphene (SLG) at $T=310$ K. It increases to ~1000 W/mK for graphite films with the thickness of 8 nm. It was also found that the suppression of thermal conductivity in encased graphene, as compared to bulk graphite, was stronger at low temperatures where $K$ was proportional to $T^\beta$ with $1.5<\beta<2$ [64]. Thermal conduction in encased FLG was limited by the rough boundary scattering and disorder penetration through graphene.

**IV. Phonon Spectra in Graphene, Few-Layer Graphene and Graphene Nanoribbons**

Intriguing thermal and electrical properties of graphene, FLG [16, 35–38, 46-48] and graphene nanoribbons (GNRs) [65-67] stimulate investigations of phonon energy spectra in these materials and structures [68-82]. The phonon energy spectrum is important for determining the sound velocity, phonon density of states, phonon-phonon or electron-phonon scattering rates, lattice heat capacity, as well as the phonon thermal conductivity. The optical phonon properties manifest themselves in Raman measurements. The number of graphene layers, their quality and stacking order can be clearly distinguished using the Raman spectroscopy [38, 41-42, 83-84]. For these reasons, significant efforts have been made to accurately determine the phonon energy dispersion in graphite [68-71], graphene [38, 62, 72-77, 82], GNRs [78-81, 85], and to reveal specific features of their phonon modes.

The phonon dispersion in graphite along $\Gamma-M-K-\Gamma$ directions (see figure 4(a), where the graphene Brillouin zone is shown) measured by X-ray inelastic scattering was reported in Refs. [68-69]. A number of research groups calculated the phonon energy dispersion in graphite, graphene and GNRs using various theoretical approaches, including continuum model [80-81], Perdew-Burke-Ernzerhof generalized gradient approximation (GGA) [68, 70-71], first-order local density function approximation (LDA) [70, 72, 76], fourth- and fifth-nearest neighbor force constant (4NNFC and 5NNFC) approaches [69, 71, 77], Born-von Karman or valence force field (VFF) model of the lattice dynamics [38, 73-74, 82], utilized the Tersoff and Brenner potentials [75] or Tersoff and Lennard-Jones potentials [62-63]. These models (with exception of GGA and LDA models) are based on different sets of the fitting parameters, which are usually determined from comparison with the experimental phonon dispersion, thermal expansion or heat capacity [68-69,86].

[Figure 4]

The number of parameters in the theoretical models depends on the model specifics and the number of the considered atomic neighbors. The number of the parameters varies from 5 [71] to 23 [77]. For example, VFF model developed for graphene by Nika et al. in Ref. [82] used only six parameters. In this model, all interatomic forces are resolved into bond-stretching and bond-bending forces [82, 87-89]. This model takes into account stretching and bending interactions with two in-plane and two out-of-plane atomic neighbors as well as doubled stretching-stretching interactions with the nearest in-plane neighbors [82]. The honeycomb crystal lattice of graphene utilized in this model is presented in figure 4(b). The rhombic unit cell of graphene, shown as a dashed region, contains two atoms and is defined by two basis vectors $\vec{a}_1 = a(3,\sqrt{3})/2$, and $\vec{a}_2 = a(3,-\sqrt{3})/2$, where $a = 0.142$ nm is the distance between two nearest carbon atoms. The six phonon polarization branches $s = 1,…, 6$ in SLG are shown in figure 5. These branches are (i) out-of-plane acoustic (ZA) and out-of-plane optical (ZO) phonons with the displacement vector along the Z axis; (ii) transverse acoustic (TA) and transverse optical (TO) phonons, which corresponds to the transverse vibrations within the graphene plane; (iii) longitudinal acoustic (LA) and longitudinal optical (LO) phonons, which corresponds to the longitudinal vibrations within the graphene plane.

[Figure 5]

Although various theoretical models are in qualitative agreement with each other, they predict substantially different phonon frequencies at the $\Gamma$, M or K points of the Brillouin zone. Moreover, some of the models give the same frequencies for the LO - LA phonons [71-72,75] and ZO - TA phonons [69-70,73,82] at the M point while the rest of the models predict different frequencies for these phonons at the M point [68,74,76]. The comparison between phonon frequencies at the high-symmetry points of the Brillouin zone is presented in Tables I and II. The discrepancy in the calculated phonon dispersion can easily translate into differences in the predicted thermal conductivity values. Specifically, the relative contribution of the LA, TA, and ZA phonons to heat transport may vary in a wide range depending on the specifics of the phonon dispersion used.

[Table I]

[Table II]

The unit cell of the *n*-layer graphene contains $2 \cdot n$ atoms, therefore $6 \cdot n$ quantized phonon branches appear in *n*-layer graphene. In figure 6(a-b) we show the phonon dispersions in bilayer graphene. Weak van der Waals interaction between monolayers leads to the coupling of long wavelength phonons only and quantization of the low-energy part of the spectrum with *q<0.1q$_{max}$* for LA, TA, LO, TO and ZO phonons and with *q<0.4q$_{max}$* for ZA phonons (see figure 6(b)). The modification of the phonon energy spectrum in *n*-layer graphene as compared with that in single layer graphene results in a substantial change of the three-phonon scattering rates and a reduction of the intrinsic thermal conductivity in *n*-layer graphene [38,62-63].

[Figure 6]

V.  **Specifics of the Acoustic Phonon Transport in Two-Dimensional Crystals**

We now address in more detail some specifics of the acoustic phonon transport in 2D systems. Investigation of the heat conduction in graphene [35-36] and CNTs [90] raised the issue of ambiguity in the definition of the intrinsic thermal conductivity for 2D and 1D crystal lattices. It was theoretically shown that the intrinsic thermal conductivity limited by the crystal anharmonicity has a finite value in 3D bulk crystals [12, 59]. However, many theoretical models predict that the intrinsic thermal conductivity reveals a logarithmic divergence in strictly 2D systems, $K \sim ln(N)$, and the power-law divergence in 1D systems, $K \sim N^{\alpha}$, with the number of atoms $N$ ($0<\alpha<1$) [12, 17, 59, 90-94]. The logarithmic divergence can be removed by introduction of the *extrinsic* scattering mechanisms such as scattering from defects or coupling to the substrate [59]. Alternatively, one can define the *intrinsic* thermal conductivity of a 2D crystal for a given size of the crystal.

Graphene is not an ideal 2D crystal, considered in most of the theoretical works, since graphene atoms vibrate in three directions. Nevertheless, the intrinsic graphene thermal conductivity strongly depends on the graphene sheet size due to weak scattering of the low-energy phonons by other phonons in the system. Therefore, the phonon boundary scattering is an important mechanism for phonon relaxation in graphene. Different studies [95-97] also suggested that an accurate accounting of the higher-order anharmonic processes, i.e. above three-phonon Umklapp scattering, and inclusion of the normal phonon processes into consideration allow one to limit the low-energy phonon MFP. The normal phonon processes do not contribute directly to thermal resistance but affect the phonon mode distribution [62, 97-98]. However, even these studies found that the graphene sample has to be very large (>10 μm) to obtain the size-independent thermal conductivity.

The specific phonon transport in the quasi - 2D system such as graphene can be illustrated with an expression derived by Klemens specifically for graphene [25, 99]. In the framework of the BTE approach and the RTA, the intrinsic Umklapp-limited thermal conductivity of graphene can be written as [25, 99]:

$$K = \frac{\rho_m}{2\pi\gamma^2} \frac{\bar{\upsilon}^4}{f_m T} \ln(\frac{f_m}{f_B}). \qquad (7)$$

Here, $f_m$ is the upper limit of the phonon frequencies defined by the phonon dispersion, $\bar{\upsilon}$ is the average phonon group velocity, $f_B = \left(M\bar{\upsilon}^3 f_m / 4\pi\gamma^2 k_B TL\right)^{1/2}$ is the size-dependent low-bound cut-off frequency for acoustic phonons, introduced by limiting the phonon MFP with the graphene layer size $L$.

In Ref. [100] we improved equation (7) by taking into account the actual maximum phonon frequencies and Gruneisen parameters $\gamma_s$ (s=TA, LA) determined separately for LA and TA phonon branches. The Gruneisen parameters were computed by averaging the phonon mode-dependent $\gamma_s(\vec{q})$ for all relevant phonons (here $\vec{q}$ is the wave vector):

$$K = \frac{1}{4\pi k_B T^2 h} \sum_{s=TA,LA} \int_{q_{min}}^{q_{max}} \{[\hbar\omega_s(q)\frac{d\omega_s(q)}{dq}]^2 \tau_{U,s}^K(q) \frac{exp[\hbar\omega_s(q)/k_B T]}{[exp[\hbar\omega_s(q)/k_B T]-1]^2} q\}dq. \qquad (8)$$

Here $\hbar\omega_s(q)$ is the phonon energy, $h = 0.335$ nm is the graphene layer thickness and $\tau_{U,s}^K(q)$ is the three-phonon mode-dependent Umklapp relaxation time, which was derived using an expression from Refs. [25-26] but introducing separate life times for *LA* and *TA* phonons:

$$\tau_{U,s}^K = \frac{1}{\gamma_s^2} \frac{M\bar{v}_s^2}{k_B T} \frac{\omega_{s,\max}}{\omega^2}, \tag{9}$$

where $s=TA, LA$, $\bar{v}_s$ is the average phonon velocity for a given branch, $\omega_{s,\max} = \omega(q_{\max})$ is the maximum cut-off frequency for a given branch and $M$ is the mass of a graphene unit cell. In Refs. [25, 99-100] the contribution of ZA phonons to thermal transport has been neglected because of their low group velocity and large Gruneisen parameter $\gamma_{ZA}$ [70, 100]. Equation (9) can be used to calculate thermal conductivity with the actual dependence of the phonon frequency $\omega_s(q)$ and the phonon velocity $d\omega_s(q)/dq$ on the phonon wave number. To simplify the model, one can use the liner dispersion $\omega_s(q) = \bar{v}_s q$ and re-write it as:

$$K_U = \frac{\hbar^2}{4\pi k_B T^2 h} \sum_{s=TA,LA} \int_{\omega_{\min}}^{\omega_{\max}} \{\omega^3 \tau_{U,s}^K(\omega) \frac{\exp[\hbar\omega/kT]}{[\exp[\hbar\omega/kT]-1]^2}\} d\omega. \tag{10}$$

Substituting equation (9) to equation (10) and performing integration one obtains

$$K_U = \frac{M}{4\pi T h} \sum_{s=TA,LA} \frac{\omega_{s,\max} \bar{v}_s^2}{\gamma_s^2} F(\omega_{s,\min}, \omega_{s,\max}), \tag{11}$$

where

$$F(\omega_{s,\min}, \omega_{s,\max}) = \int_{\hbar\omega_{s,\min}/k_B T}^{\hbar\omega_{s,\max}/k_B T} \xi \frac{\exp(\xi)}{[\exp(\xi)-1]^2} d\xi =$$
$$[\ln\{\exp(\xi)-1\} + \frac{\xi}{1-\exp(\xi)} - \xi]\Big|_{\hbar\omega_{s,\min}/k_B T}^{\hbar\omega_{s,\max}/k_B T}. \tag{12}$$

In the above equation, $\xi = \hbar\omega/k_B T$, and the upper cut-off frequencies $\omega_{s,\max}$ are defined from the actual phonon dispersion in graphene (see figure 5): $\omega_{LA,\max} = 2\pi f_{LA,\max}(\Gamma K) = 241$ rad/ps, $\omega_{TA,\max} = 2\pi f_{TA,\max}(\Gamma K) = 180$ rad/ps.

The integrand in equation (12) can be further simplified near RT when $\hbar\omega_{s,\max} > k_B T$, and it can be expressed as

$$F(\omega_{s,\min}) \approx -ln\{|exp(\hbar\omega_{s,\min}/k_BT)-1|\} + \frac{\hbar\omega_{s,\min}}{k_BT}\frac{exp(\hbar\omega_{s,\min}/k_BT)}{exp(\hbar\omega_{s,\min}/k_BT)-1}. \quad (13)$$

There is a clear difference between the heat transport in basal planes of bulk graphite and in single layer graphene [25, 99]. In the former, the heat transport is approximately two-dimensional only up to some lower-bound cut-off frequency $\omega_{\min}$. Below $\omega_{\min}$ there appears to be strong coupling with the cross-plane phonon modes and heat starts to propagate in all directions, which reduces the contributions of these low-energy modes to heat transport along basal planes to negligible values. In bulk graphite, there is a physically reasonable reference point for the on-set of the cross-plane coupling, which is the ZO' phonon branch near ~4 THz observed in the spectrum of bulk graphite [25, 101]. The presence of the ZO' branch and corresponding $\omega_{\min} = \omega_{ZO'}(q=0)$ allows one to avoid the logarithmic divergence in the Umklapp-limited thermal conductivity integral [see equations (10–13)] and calculate it without considering other scattering mechanisms.

The physics of heat conduction is principally different in graphene where the phonon transport is 2D all the way to zero phonon frequency $\omega(q=0)=0$. There is no onset of the cross-plane heat transport at the long-wavelength limit in the system, which consists of only one atomic plane. This is no *ZO'* branch in the phonon dispersion of graphene (see figure 5). Therefore the lower-bound cut-off frequencies $\omega_{s,\min}$ for each *s* are determined from the condition that the phonon MFP cannot exceed the physical size *L* of the flake, i.e.

$$\omega_{s,\min} = \frac{\bar{\upsilon}_s}{\gamma_s}\sqrt{\frac{M\bar{\upsilon}_s}{k_BT}\frac{\omega_{s,\max}}{L}}. \quad (14)$$

We would like to emphasize here that using size-independent graphite $\omega_{\min}$ for SLG or FLG (as has been proposed in Ref. [102]) is without scientific merit and leads to an erroneous calculation of thermal conductivity, as described in detail in Ref. [103]. Equations (12-14) constitute a simple analytical model for the calculation of the thermal conductivity of the graphene layer, which retains such important features of graphene phonon spectra as different $\bar{\upsilon}_s$ and $\gamma_s$ for *LA* and *TA* branches. The model also reflects the two-dimensional nature of heat transport in graphene all the way down to zero phonon frequency.

In figure 7, we present the dependence of thermal conductivity of graphene on the dimension of the flake *L*. The data is presented for the averaged values of the Gruneisen parameters $\gamma_{LA}$=1.8 and $\gamma_{TA}$=0.75 obtained from *ab initio* calculations, as well as for several other close sets of $\gamma_{LA,TA}$ to illustrate the sensitivity of the result to the Gruneisen parameters. For small graphene flakes, the *K* dependence on *L* is rather strong. It weakens for flakes with *L*≥10 μm. The calculated values are in good agreement with available experimental data for suspended exfoliated [35-36] and CVD graphene [49-50]. The horizontal dashed line indicates the experimental thermal conductivity for bulk graphite, which is exceeded by graphene's thermal conductivity at smaller *L*. Thermal conductivity, presented in figure 7, is an *intrinsic* quantity limited by the three-phonon Umklapp scattering only. But it is determined for a specific graphene flake size since *L* defines the lower-bound (long-wavelength) cut-off frequency in Umklapp scattering through equation (14). In experiments, thermal conductivity is also limited by defect scattering. When the size of the flake becomes very large with many polycrystalline grains, the scattering on their boundaries will also lead to phonon relaxation. The latter can be included in our model through adjustment of *L*. The extrinsic phonon scattering mechanisms or high-order phonon-phonon scatterings prevent indefinite growth of thermal conductivity of graphene with *L*.

[Figure 7]

**VI. The Q-Space Diagram Theory of Phonon Transport in Graphene**

The simple models described in the previous section are based on the Klemens-like expressions for the relaxation time (see equation (9)). Therefore they do not take into account all peculiarities of the 2D three-phonon Umklapp processes in SLG or FLG, which are important for the accurate description of thermal transport. There are two types of the three-phonon Umklapp scattering processes [26]. The first type is the scattering when a phonon with the wave vector $\vec{q}(\omega)$ absorbs another phonon from the heat flux with the wave vector $\vec{q}'(\omega')$, i.e. the phonon leaves the state $\vec{q}$. For this type of scattering processes the momentum and energy conservation laws are written as:

$$\begin{aligned}\vec{q}(\omega)+\vec{q}'(\omega')=\vec{b}_i+\vec{q}''(\omega''), \; i=1,2,3 \\ \omega+\omega'=\omega''\end{aligned} \quad (15)$$

The processes of the second type are those when the phonons $\vec{q}(\omega)$ of the heat flux decay into two phonons with the wave vectors $\vec{q}'(\omega')$ and $\vec{q}''(\omega'')$, i.e. leaves the state $\vec{q}(\omega)$, or, alternatively, two phonons $\vec{q}'(\omega')$ and $\vec{q}''(\omega'')$ merge together forming a phonon with the wave vector $\vec{q}(\omega)$, which correspond to the phonon coming to the state $\vec{q}(\omega)$. The conservation laws for this type are given by:

$$\vec{q}(\omega) + \vec{b}_i = \vec{q}'(\omega') + \vec{q}''(\omega''), \quad i = 4,5,6$$
$$\omega = \omega' + \omega'', \tag{16}$$

In equations (15-16) $\vec{b}_i = \vec{\Gamma\Gamma}_i, i = 1,2,...,6$ is one of the vectors of the reciprocal lattice (see figure 4(a)).

Calculations of the thermal conductivity in graphene taking into account all possible three-phonon Umklapp processes allowed by the equations (15-16) and actual phonon dispersions were carried out for the first time in Ref. [82]. For each phonon mode ($q_i$, $s$), were found all pairs of the phonon modes ($\vec{q}'$, $s'$) and ($\vec{q}''$, $s''$) such that the conditions of equations (15-16) are met. As a result, in ($\vec{q}'$)-space were constructed the *phase diagrams* for all allowed three-phonon transitions [82]. Using the long-wave approximation (LWA) for a matrix element of the three-phonon interaction authors of Ref. [82] obtained for the Umklapp scattering rates

$$\frac{1}{\tau_U^{(I),(II)}(s,\vec{q})} = \frac{\hbar \gamma_s^2(\vec{q})}{3\pi\rho\upsilon_s^2(\vec{q})} \sum_{s's'';\vec{b}_i} \iint \omega_s(\vec{q})\omega'_{s'}(\vec{q}')\omega''_{s''}(\vec{q}'') \times$$
$$\{N_0[\omega'_{s'}(\vec{q}')] \mp N_0[\omega''_{s''}(\vec{q}'')] + \frac{1}{2} \mp \frac{1}{2}\} \times \delta[\omega_s(\vec{q}) \pm \omega'_{s'}(\vec{q}') - \omega''_{s''}(\vec{q}'')] dq'_l dq'_\perp. \tag{17}$$

Here $q'_l$ and $q'_\perp$ are the components of the vector $\vec{q}'$ parallel or perpendicular to the lines defined by equations (15-16), correspondingly, $\gamma_s(\vec{q})$ is the mode-dependent Gruneisen parameter, which is determined for each phonon wave vector and polarization branch and $\rho$ is the surface mass density. In equation (17) the upper signs correspond to the processes of the first type while the lower signs correspond to those of the second type. The integrals for $q_l, q_\perp$ are taken along and perpendicular to the curve segments, correspondingly, where the conditions of equations (15-16) are met. Integrating along $q_\perp$ in equation (17) one can obtain the line integral

$$\frac{1}{\tau_U^{(I),(II)}(s,\vec{q})} = \frac{\hbar \gamma_s^2(\vec{q}) \omega_s(\vec{q})}{3\pi \rho v_s^2(\vec{q})} \sum_{s's'';b} \int_l \frac{\pm(\omega_{s''}'' - \omega_s)\omega_{s''}''}{v_{\perp,s'}(\omega_{s'}')}(N_0' \mp N_0'' + \frac{1}{2} \mp \frac{1}{2})dq_l'. \qquad (18)$$

The phonon scattering on the rough edges of graphene can be evaluated using equation (4). The total phonon relaxation rate is given by:

$$\frac{1}{\tau_{tot}(s,q)} = \frac{1}{\tau_U(s,q)} + \frac{1}{\tau_B(s,q)}. \qquad (19)$$

The sensitivity of the room temperature thermal conductivity, calculated using equations (17-19), to the value of the specular parameter of phonon boundary scattering is illustrated in figure 8. The data is presented for different sizes (widths) of the graphene flakes. The experimental data points for suspended exfoliated [35-36] and CVD [49-50] graphene are also shown for comparison. Strong dependence of graphene thermal conductivity on tensile strain, flake size, Van-der Vaals bond strength as well as concentration of lattice defects, vacancies and wrinkles was theoretically predicted in Refs. [104-111]. Table III provides representative experimental and theoretical data for the suspended and supported graphene.

[Figure 8]

**VII. Thermal Conductivity of Graphene Nanoribbons**

Measurements of thermal properties of graphene stimulated a surge of interest in theoretical and experimental studies of heat conduction in graphene nanoribbons [65-67, 85, 113-121]. It is important to understand how lateral sizes affect the phonon transport properties from both fundamental science and practical applications point of view. In the last few years a number of theoretical works investigated phonon transport and heat conduction in graphene nanoribbons with various lengths, widths, edge roughness and defect concentrations. The authors used MD simulations [65-67, 113-116], nonequilibrium Green's function method [117-119] and BTE approaches [85, 120].

Keblinsky and co-workers [65] found from the MD study that the thermal conductivity of graphene is $K \approx 8000 - 10000$ W/mK at RT for the square graphene sheet. The $K$ value was size independent for $L>5$ nm [65]. For the ribbons with fixed $L=10$ nm and width $W$ varying from 1 to 10 nm, *K* increased from ~1000 W/mK to 7000 W/mK. The thermal conductivity in GNR with rough edges can be suppressed by orders of magnitude as compared to that in GNR with perfect edges [65, 67]. The isotopic superlattice modulation of GNR or defects of crystal lattices also significantly decreases the thermal conductivity [118-119]. The uniaxial stretching applied in the longitudinal direction enhances the low-temperature thermal conductance for the 5 nm arm-chair or zigzag GNR up to 36 % due to the stretching-induced convergence of phonon spectra to the low-frequency region [117]. Aksamija and Knezevic [85] calculated the dependence of the thermal conductivity of GNR with the width 5 nm and RMS edge roughness $\Delta = 1$ nm on temperature. The thermal conductivity was calculated taking into account the three-phonon Umklapp, mass-defect and rough edge scatterings [85]. The authors obtained RT thermal conductivity $K \sim 5500$ W/mK for the graphene nanoribbon. The study of the nonlinear thermal transport in rectangular and triangular GNRs under the large temperature biases was reported in Ref. [121]. The authors found that in short (~6 nm) rectangular GNRs, the negative differential thermal conductance exists in a certain range of the applied temperature difference. As the length of the rectangular GNR increases the effect weakens. A computational study reported in Ref. [122] predicted that the combined effects of the edge roughness and local defects play a dominant role in determining the thermal transport properties of zigzag GNRs. The experimental data on thermal transport in GNRs is very limited. In Ref. [123] the authors used an electrical self-heating methods and extracted the thermal conductivity of sub 20-nm GNRs to be more than 1000 W/mK at 700 – 800 K. A similar experimental method but with more accurate account of GNRs thermal coupling to the substrate has been used in Ref. [124]. Pop and co-workers [124] found substantially lower values of thermal conductivity of ~ 80 – 150 W/mK at RT. Wang et al. [125] employed equilibrium molecular dynamic for investigation of the thermal conductivity in graphene nanoribbons with zigzag- and armchair-edges and revealed the importance of edges to thermal conductivity in GNRs. The calculated and measured data for the thermal conductivity of graphene nanoribbons are summarized in Table IV.

[Table IV]

VIII. Conclusions

We reviewed theoretical and experimental results pertinent to two-dimensional phonon transport in graphene. Phonons are the dominant heat carriers in the non-gated graphene samples near the room temperature. The unique nature of 2D phonons, revealed in very large phonon MFP and peculiarities of the density of states, translates to unusual heat conduction properties of graphene and related materials. Recent computational studies suggest that the thermal conductivity of graphene depends strongly on the concentration of defects, strain distribution, wrinkles, sample size and geometry. The revealed dependence can account for portion of the data scatter in reported experimental studies. Investigation of the physics of 2D phonons in graphene can shed light on the thermal energy transfer in low-dimensional systems. The results presented in this chapter are important for the proposed electronic and optoelectronic applications of graphene, and can lead to new methods of heat removal and thermal management.

*Acknowledgements*

This work was supported, in part, by the National Science Foundation (NSF) projects US EECS-1128304, EECS-1124733 and EECS-1102074, by the US Office of Naval Research (ONR) through award N00014-10-1-0224, Semiconductor Research Corporation (SRC) and Defense Advanced Research Project Agency (DARPA) through FCRP Center on Functional Engineered Nano Architectonics (FENA), and DARPA-DMEA under agreement H94003-10-2-1003. DLN acknowledges the financial support through the Moldova State Project No. 11.817.05.10F and 12.819.05.18F.

**Table I: Energies of ZO and LO Phonons at Γ Point in Graphite and Graphene**

| Sample | $\Gamma_{ZO}$ (cm$^{-1}$) | $\Gamma_{LO}$ (cm$^{-1}$) | Comments | Refs |
|---|---|---|---|---|
| graphite | --- | 1583[a] | experiment: X-ray scattering | [a]68 [b]69 [c]70 [d]86 [e]71 [f]72 [g]76 |
| graphite | --- | 1581[b] | experiment: X-ray scattering | |
| graphite | 899[c] | 1593[c] | theory: LDA | |
| graphite | ~820[a], 879[c], 881[c] | 1559[c], 1561[c], 1581-1582[a] | theory: GGA | |
| graphite | 868[b] | 1577[b] | theory: 5NNFC | |
| graphite | ~920[d] | ~1610[d] | theory: six-parameter force constant model | |
| graphene | 879[c], 881[c], 884[e] | 1554[c], 1559[c], 1569[e] | theory: GGA | |
| graphene | 890[g], 896[g], ~900[f] | 1586[f], 1595[g], 1597[g] | theory: LDA | |
| graphene | 893 | 1581 | theory: Born-von Karman | 73 |
| graphene | 889[h], 883.5[i] | 1588[h], 1555[i] | theory: VFF model | [h]74 [i]82 |
| graphene | ~1300 | ~1685 | theory: optimized Tersoff | 75 |
| | ~1165 | ~1765 | theory: optimized Brenner | |

**Table II. Phonon Energies at K and M Points in Graphite and Graphene**

| Sample | $K_{ZA}$ (cm$^{-1}$) | $K_{TA}$ (cm$^{-1}$) | $K_{LA}$ (cm$^{-1}$) | Comments | Refs |
|---|---|---|---|---|---|
| graphite | --- | --- | 1194[a] | experiment: X-ray scattering; $\omega_{LO}(M) > \omega_{LA}(M)$; | [a]68 |
| graphite | 542[b] | 1007[b] | 1218[b] | experiment: X-ray scattering; $\omega_{LO}(M) > \omega_{LA}(M)$; $\omega_{ZO}(M) \approx \omega_{TA}(M)$ | [b]69 |
| graphite | --- | --- | --- | experiment: high-resolution electron-energy-loss spectroscopy; $\omega_{LO}(M) > \omega_{LA}(M)$; $\omega_{ZO}(M) < \omega_{TA}(M)$ | 86 |
| graphite | 540[c] | 1009[c] | 1239[c] | theory: LDA; $\omega_{LO}(M) > \omega_{LA}(M)$; $\omega_{ZO}(M) \approx \omega_{TA}(M)$ | [c]70 [d]71 |
| graphite | 534[c], 540[c] | ~960[a], 998[c], 999[c] | 1220[a], 1216[c], 1218[c] | theory: GGA; [a,c]$\omega_{LO}(M) > \omega_{LA}(M)$; [c]$\omega_{ZO}(M) \approx \omega_{TA}(M)$ | |
| graphite | 542[b] | 1007[b] | 1218[b] | theory: 5NNFC; $\omega_{LO}(M) > \omega_{LA}(M)$; $\omega_{ZO}(M) \approx \omega_{TA}(M)$ | |
| graphene | 535[c], 539[d] | 997[c], 1004[d] | 1213[c], 1221[d] | theory: GGA; [c]$\omega_{LO}(M) > \omega_{LA}(M)$; [d]$\omega_{LO}(M) \approx \omega_{LA}(M)$; [c,d]$\omega_{ZO}(M) \approx \omega_{TA}(M)$ | |
| graphene | ~520[e,f] | ~990[f] ~1000[e] | ~1250[f], ~1220[e] | theory: LDA; [e]$\omega_{LO}(M) \approx \omega_{LA}(M)$; [e]$\omega_{ZO}(M) \approx$ $\approx \omega_{ZA}(M) \ll \omega_{TA}(M)$ [f]$\omega_{LO}(M) > \omega_{LA}(M)$; [f]$\omega_{ZO}(M) > \omega_{ZA}(M)$; | [e]72 [f]76 |
| graphene | 495 | 1028 | 1199 | theory: Born-von Karman model; $\omega_{LO}(M) > \omega_{LA}(M)$; $\omega_{ZO}(M) \approx \omega_{TA}(M)$ | 73 |
| graphene | 544[g], 532[h] | 1110[g], 957[h] | 1177[g], 1267[h] | theory: VFF model; [g,h]$\omega_{LO}(M) > \omega_{LA}(M)$; [g]$\omega_{ZO}(M) < \omega_{TA}(M)$; [h]$\omega_{ZO}(M) \approx \omega_{TA}(M)$ | [g]74 [h]82 |

| graphene | ~635 | ~1170 | ~1170 | theory: optimized Tersoff potential; $\omega_{LO}(M) \approx \omega_{LA}(M)$; $\omega_{ZO}(M) > \omega_{TA}(M)$ | 75 |
| --- | --- | --- | --- | --- | --- |
| | ~585 | ~1010 | ~1240 | theory: optimized Brenner potential; $\omega_{LO}(M) > \omega_{LA}(M)$; $\omega_{ZO}(M) > \omega_{TA}(M)$ | |

**Table III: Thermal conductivity of graphene and few-layer graphene**

| Sample | K (W/mK) | Method | Comments | Refs |
|---|---|---|---|---|
| graphene | ~2000 – 5000 | Raman optothermal | suspended; exfoliated | 35,36 |
| FLG | 1300 - 2800 | Raman optothermal | suspended; exfoliated; n=2-4 | 38 |
| graphene | ~2500 | Raman optothermal | suspended; CVD | 49 |
| graphene | ~1500-5000 | Raman optothermal | suspended; CVD | 50 |
| graphene | 600 | Raman optothermal | suspended; exfoliated; T ~ 660 K | 51 |
| graphene | 600 | Electrical | supported; exfoliated; | 57 |
| graphene | ~ 1875 at T = 420 K | Micro-Raman mapping | suspended CVD graphene membranes with and without wrinkles; wrinkles decreases the thermal conductivity by ~ 27 percents. | 107 |
| graphene | ~2430 | Theory: BTE, 3$^{rd}$-order IFCs | $K(graphene) \geq K(carbon\,nanotube)$ | 98 |
| graphene | 1000 - 8000 | Theory: BTE+RTA $\gamma_{LA}, \gamma_{TA}$ | strong size dependence | 100 |
| graphene | 2000-8000 | Theory: BTE+RTA, $\gamma_s(q)$ | strong edge, width and grunaisen parameter dependence | 82 |
| graphene | ~ 4000 | Theory: ballistic | strong width dependence | 66 |
| graphene | ~ 2900 | Theory: MD simulation | strong dependence on the vacancy concentration | 104 |
| graphene | 1500 - 3500 | Theory: BTE, 3$^{rd}$-order IFCs | strong size dependence | 108 |
| graphene | ~ 5000 | Theory: BTE + RTA | strong size and defect concentration dependence | 109 |
| graphene | ~1780 (suspended) ~ 480 (supported) | Theory: equilibrium molecular dynamics | supported on copper; strong reduction of the thermal conductivity in supported graphene | 110 |
| FLG | 1000 - 4000 | Theory: BTE+RTA, $\gamma_s(q)$ | n = 8 – 1, strong size dependence | 38 |
| FLG | 1000 - 3500 | Theory: BTE, 3$^{rd}$-order IFCs | n = 5 – 1, strong size dependence | 62 |

| FLG | 2000-3300 | Theory: BTE, $3^{rd}$-order IFCs | n = 4 – 1 | 63 |
| FLG | 580 - 880 | Theory: MD simulation | n = 5 – 1, strong dependence on the Van-der Vaals bond strength | 111 |

**Table IV: Thermal conductivity of graphene nanoribons**

| | | | | |
|---|---|---|---|---|
| FLG nanoribbon | 1100 | Electrical self-heating | supported; exfoliated; n<5 | 109 |
| FLG nanoribbon | 80 - 150 | Electrical self-heating | supported | 110 |
| GNR | 1000 - 7000 | Theory: molecular dynamics, Tersoff | strong ribbon width and edge dependence | 59 |
| GNR | ~ 5500 | Theory: BTE + RTA | GNR with width of 5 μm; strong dependence on the edge roughness | 81 |
| GNR | 400 - 600 | Theory: equilibrium molecular dynamics | GNR with width of 4 nm; strong dependence on the GNR edge and thickness | 125 |

**Figure Captions**

**Figure 1**: (a) Schematic of the experimental set up with the excitation laser light focused on graphene suspended across a trench in Si wafer. Laser power absorbed in graphene induces a local hot spot and generates heat wave propagating toward the heat sinks. (b) Illustration of the micro- and nanoscale corrugation formed in the suspended flake, which further reduce the thermal coupling to the substrate. The depicted experimental technique allows one for the steady-state non-contact direct measurement of the thermal conductivity. Figure is after Ref. [37] reproduced with permission from with permission from the Institute of Physics and Deutsche Physikalische Gesellschaft.

**Figure 2**: Illustration of optothermal micro-Raman measurement technique developed for investigation of phonon transport in graphene. (a) Schematic of the thermal conductivity measurement showing suspended FLG flakes and excitation laser light. (b) Optical microscopy images of FLG attached to metal heat sinks. (c) Colored scanning electron microscopy image of the suspended graphene flake to clarify typical structure geometry. (d) Experimental data for Raman G-peak position as a function of laser power, which determines the local temperature rise in response to the dissipated power. (e) Finite-element simulation of temperature distribution in the flake with the given geometry used to extract the thermal conductivity. Figure is after Ref. [38] reproduced with permission from the Nature Publishing Group.

**Figure 3:** Measured thermal conductivity as a function of the number of atomic planes in FLG. The dashed straight lines indicate the range of bulk graphite thermal conductivities. The blue diamonds were obtained from the first-principles theory of thermal conduction in FLG based on the actual phonon dispersion and accounting for all allowed three-phonon Umklapp scattering channels. The green triangles are Callaway–Klemens model calculations, which include extrinsic effects characteristic for thicker films. Figure is after Ref. [38] reproduced with permission from the Nature Publishing Group.

**Figure 4:** (a) Reciprocal lattice of graphene. (b) Graphene crystal lattice. The rhombic unit cell is shown as a shaded region. Figure is reproduced from Ref. [82] with permission from the American Physical Society.

**Figure 5:** Phonon frequencies $\omega_s$ in graphene calculated using the Valence Force Field model. Figure is reproduced from Ref. [37] with permission from the Institute of Physics and Deutsche Physikalische Gesellschaft.

**Figure 6:** Phonon energy spectra in bilayer graphene calculated using the valence force field model shown for (a) $\Gamma - M$ direction and (b) near the Brillouin zone center. Figure is after Ref. [38] reproduced with permission from the Nature Publishing Group.

**Figure 7:** Calculated room temperature thermal conductivity of graphene as a function of the lateral size for several values of the Gruneisen parameter. Experimental data points from Refs. [35-36] (circle), [49] (square) and [50] (rhomb) are shown for comparison. Figure is after Ref. [112] reproduced with permission from the Institute of Physics (IOP).

**Figure 8:** Calculated room temperature thermal conductivity of suspended graphene as a function of the specularity parameter $p$ for the phonon scattering from the flake edges. Note a strong dependence on the size of the graphene flakes. Experimental data points from Refs. [35-36] (circle), [49] (square) and [50] (rhomb) are shown for comparison. Figure is after Ref. [112] reproduced with permission from the Institute of Physics (IOP).

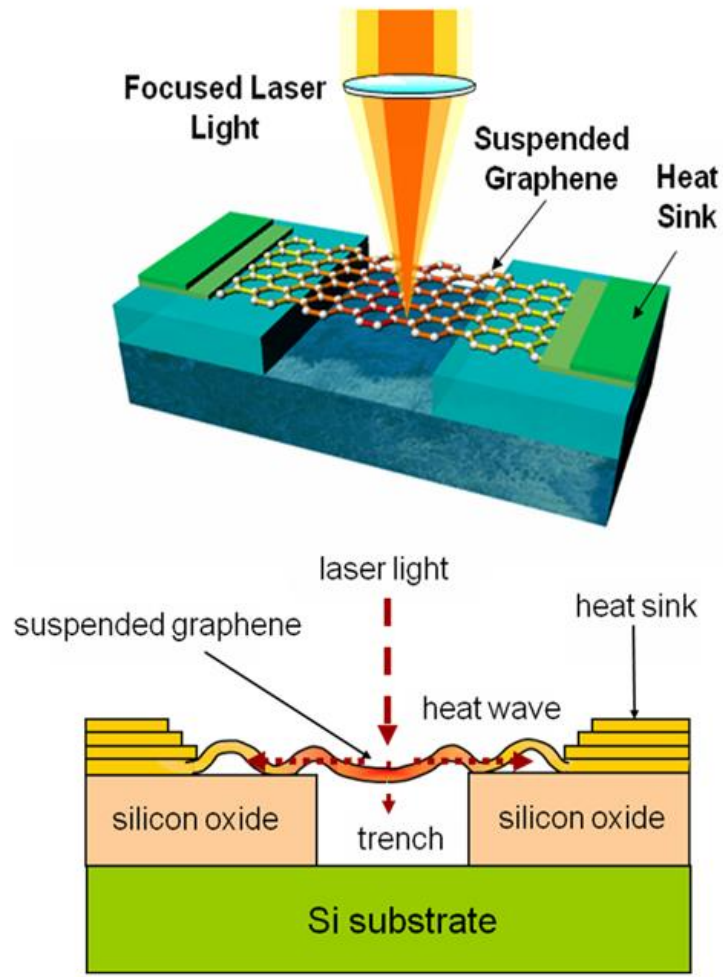

Figure 1 of 8: A.A. Balandin and D.L. Nika

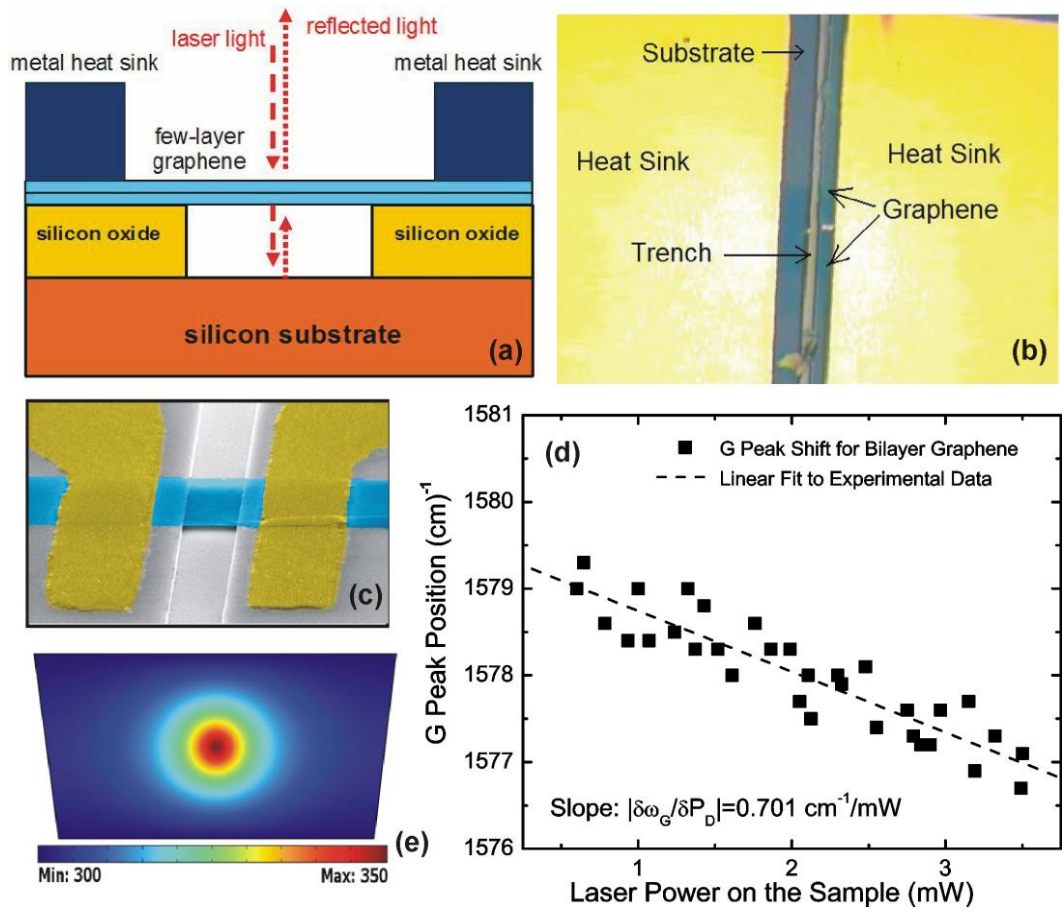

Figure 2 of 8: A.A. Balandin and D.L. Nika

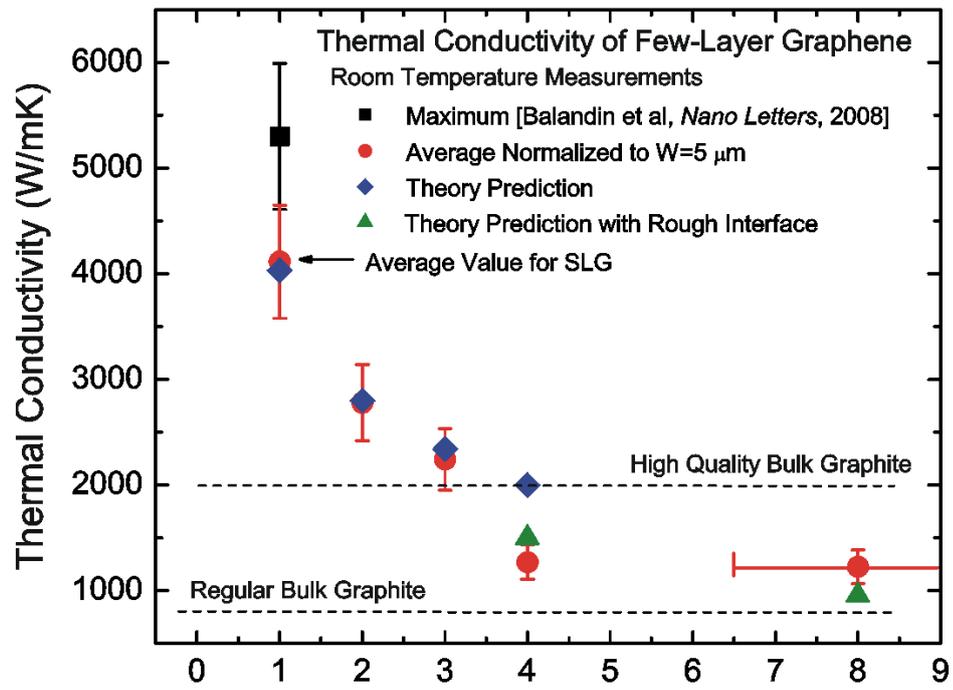

Figure 3 of 8: A.A. Balandin and D.L. Nika

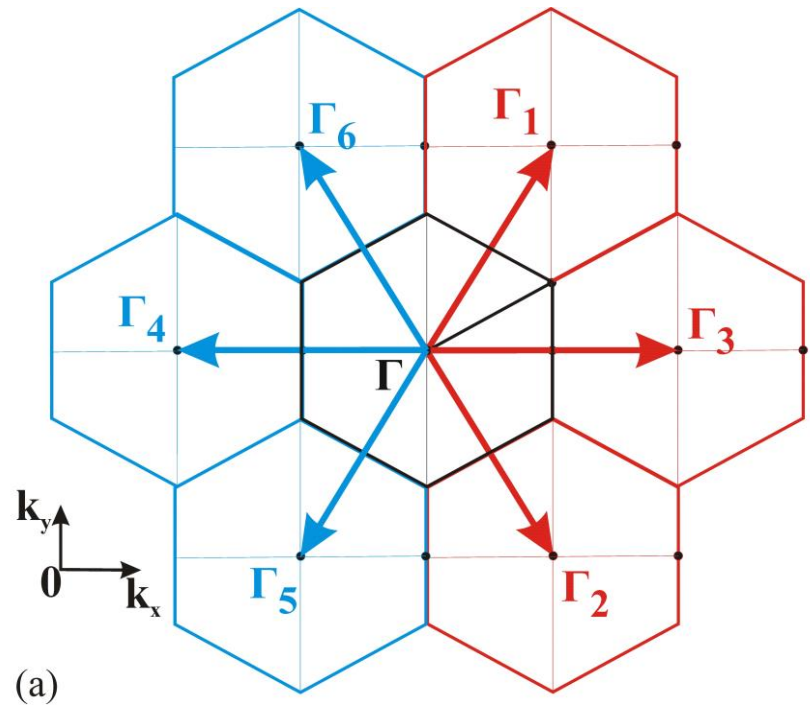

(a)

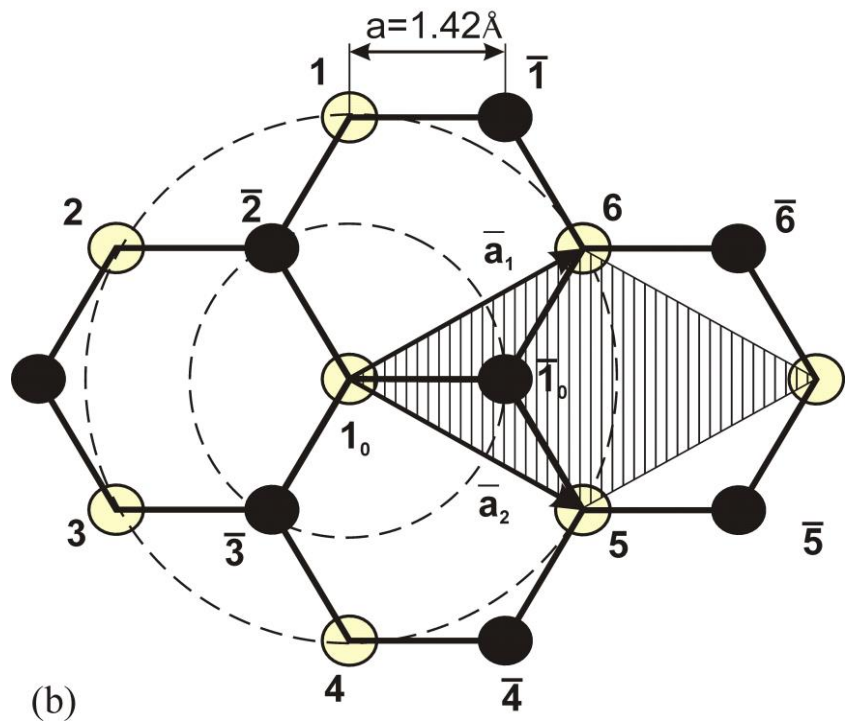

(b)

Figure 4 of 8: A.A. Balandin and D.L. Nika

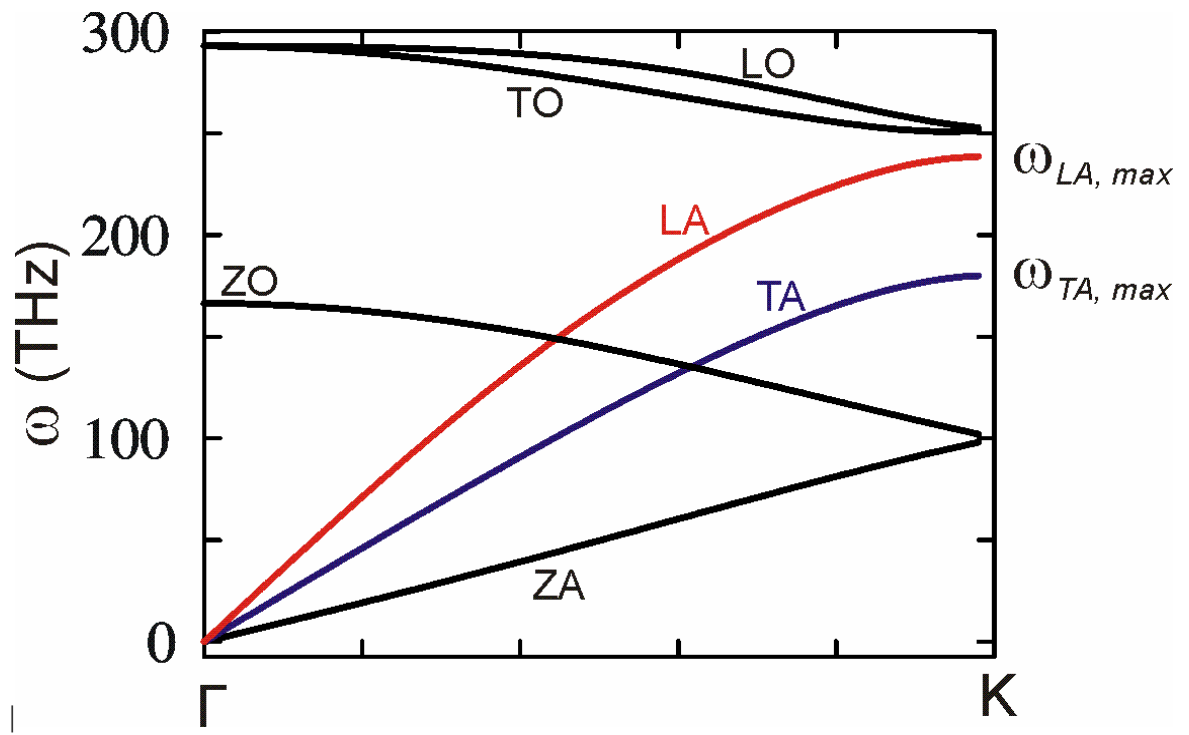

Figure 5 of 8: A.A. Balandin and D.L. Nika

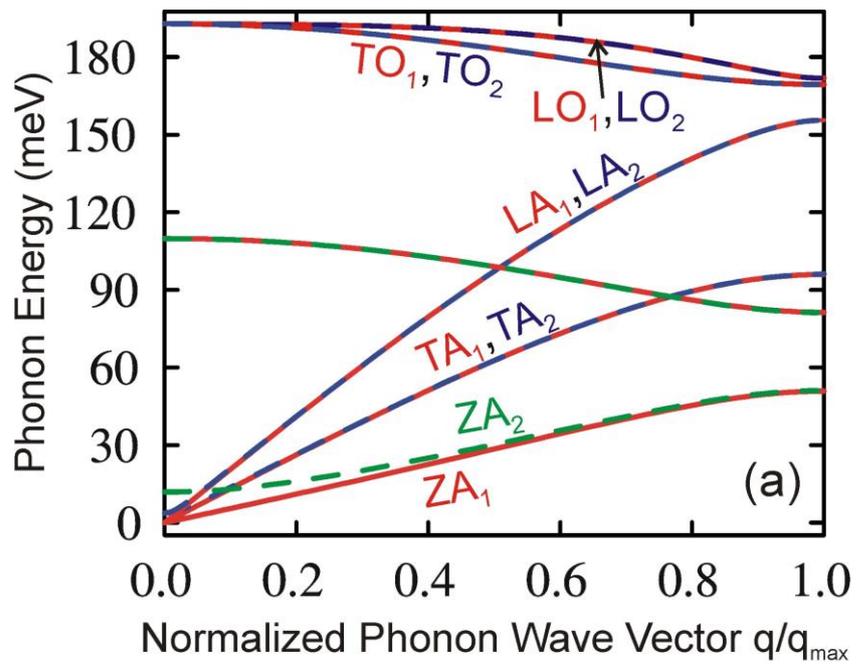

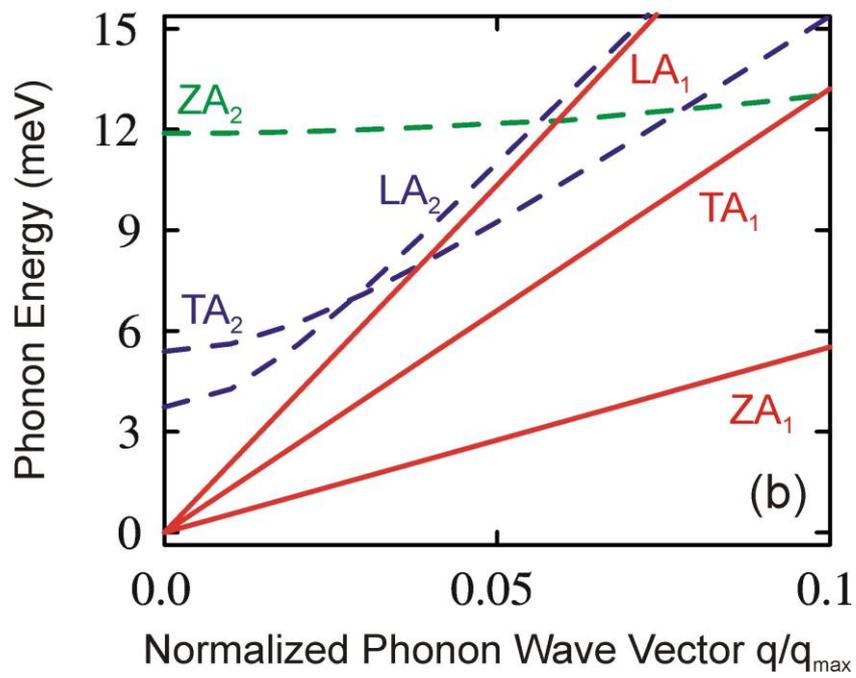

Figure 6 of 8: A.A. Balandin and D.L. Nika

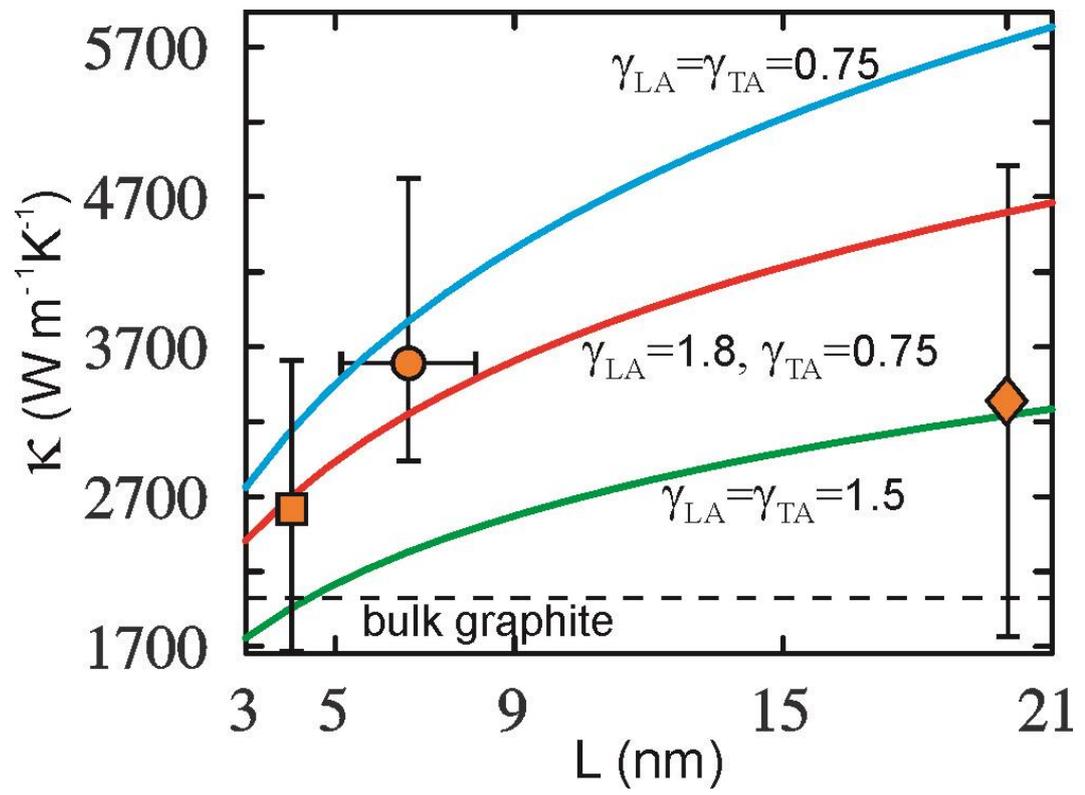

Figure 7 of 8: A.A. Balandin and D.L. Nika

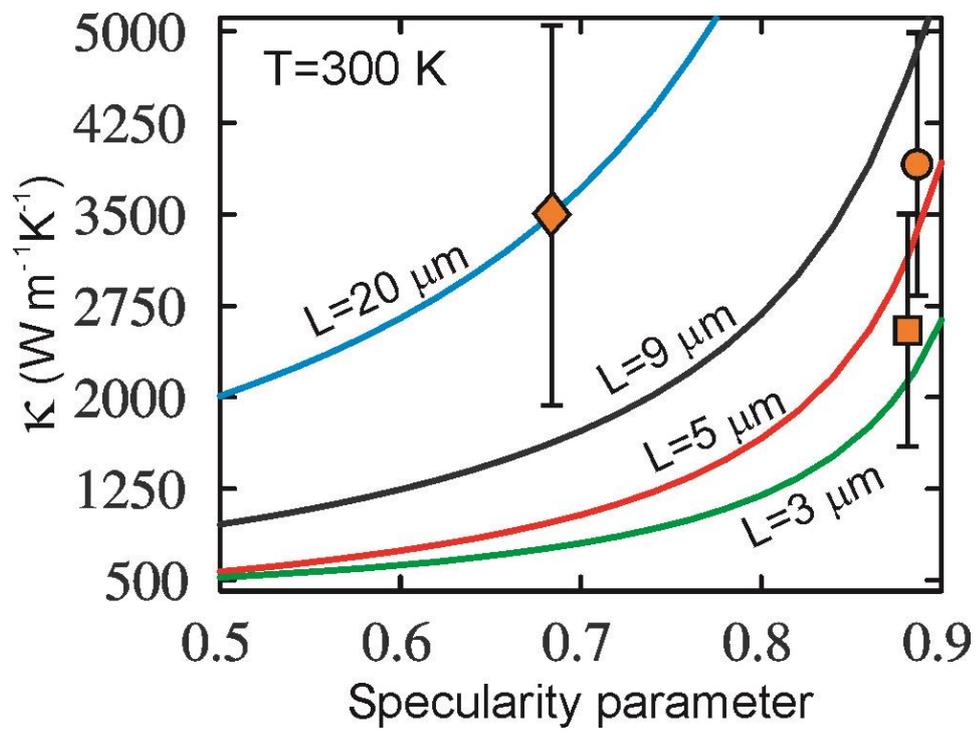

Figure 8 of 8: A.A. Balandin and D.L. Nika